\documentclass[authoryear]{article}

\usepackage[english]{babel}
\usepackage{fancyhdr} 
\usepackage{url} 
\usepackage{caption} 
\usepackage{subcaption} 
\usepackage{natbib} 
\usepackage{multirow} 
\usepackage{float} 
\usepackage{amssymb}
\usepackage[intlimits]{amsmath} 
\usepackage{amsthm} 
\usepackage[german]{nomencl}
\usepackage{booktabs} 
\usepackage{geometry}
\usepackage{enumitem}
\usepackage{bbm}
\usepackage{authblk}
\usepackage{graphicx}
\usepackage{tcolorbox}
\usepackage{listings}
\usepackage{xcolor}
\usepackage{bm}
\usepackage{booktabs} 
\usepackage{siunitx}

\usepackage{lscape}

\makenomenclature
\usepackage[pdfborder={000},colorlinks=true,linkcolor=blue,citecolor=blue]{hyperref}
\usepackage{listings}
\usepackage[all]{xy}

\usepackage{color}

\definecolor{b}{rgb}{0,0,.8}	%%omega-blau
\definecolor{g}{rgb}{0,.6,0}	%%Tau-grün
\definecolor{n}{rgb}{0,0,0}	%%normal-schwarz
\definecolor{h}{rgb}{0.4,0.2,0.2}	%%hint
\definecolor{v}{rgb}{0.2,0.6,0}

\DeclareMathOperator*{\argmax}{arg\,max}

\newtcolorbox[auto counter]{mybox}[1][]{
  colback=lightgray,    
  boxrule=1pt,          
  colframe=black,       
  arc=3mm,              
  auto outer arc,       
  width=\linewidth,     
  boxsep=5pt,           
  halign=left,        
  valign=center,
  title=Message Box \thetcbcounter: #1,
  fonttitle=\bfseries,
  label={box\thetcbcounter}
}

\lstdefinestyle{pythonstyle}{
  language=Python,                           % Specify the language for syntax highlighting
  basicstyle=\ttfamily\small,               % Set the basic style
  numbers=left,                             % Line numbers on the left
  numberstyle=\tiny\color{gray},            % Style for line numbers
  stepnumber=1,                             % Step between two line numbers
  numbersep=10pt,                           % Distance from line numbers to code
  tabsize=4,                                % Tab size
  extendedchars=true,                       % Allows 256 character set to be used
  breaklines=true,                          % Wrap lines
  keywordstyle=\color{darkgray}\bfseries,   % Styles for keywords (like def, if, for...)
  stringstyle=\color{black},                % Style for strings
  commentstyle=\color{black!70}\itshape,   % Style for comments
  showspaces=false,                         % Do not show spaces
  showtabs=false,                           % Do not show tabs
  xleftmargin=15pt,                         % Left margin
  framexleftmargin=15pt,                    % Extra margin on left for line numbers
  frame=tb,                                 % Frame on top and bottom
  framerule=0.5pt,                          % Width of frame lines
  rulecolor=\color{gray},                   % Color of frame lines
  backgroundcolor=\color{gray!10},          % Background color for the code
}

\begin{document}

\title{Linking microblogging sentiments to stock price movement:  \\ An application of GPT-4}

\author[$\star$]{Rick~Steinert}
%\ead{steinert@europa-uni.de}

\author{Saskia Altmann}
%\ead{shivarova@europa-uni.de}

\affil[$\star$]{Europa-Universit\"at Viadrina, Gro\ss e Scharrnstra\ss e 59, 15230 Frankfurt (Oder), Germany, steinert@europa-uni.de, ORCID: 0000-0002-1680-9811, corresponding author}
\affil[1]{Europa-Universit\"at Viadrina, Gro\ss e Scharrnstra\ss e 59, 15230 Frankfurt (Oder), Germany, altmann@europa-uni.de, ORCID: 0009-0009-7308-4835}

\maketitle

\begin{center}
\line(1,0){420}
\end{center}
%\begin{frontmatter}
\begin{abstract}
This paper investigates the potential improvement of the GPT-4 Language Learning
Model (LLM) in comparison to BERT for modeling same-day daily stock price movements of Apple
and Tesla in 2017, based on sentiment analysis of microblogging messages.
We recorded daily adjusted closing prices and translated them into up-down
movements. Sentiment for each day was extracted from messages on the Stocktwits platform using both LLMs. We develop a novel method to engineer a comprehensive prompt for contextual
sentiment analysis which unlocks the true capabilities of modern LLM. This enables us to carefully retrieve sentiments, perceived advantages
or disadvantages, and the relevance towards the analyzed company. Logistic regression is used to evaluate whether the extracted message contents reflect stock price movements.
As a result, GPT-4 exhibited substantial accuracy, outperforming BERT in five out of six months
and substantially exceeding a naive buy-and-hold strategy, reaching a peak accuracy of 71.47\% in
May. The study also highlights the importance of prompt engineering in obtaining desired outputs
from GPT-4’s contextual abilities. However, the costs of deploying GPT-4 and the need for fine-tuning prompts highlight some practical considerations for its use.

\end{abstract}
%\end{frontmatter}
\begin{center}
\line(1,0){420}
\end{center}

%\begin{keyword}

\textbf{Statement of Funding:} The authors report they have not received any financial funding.

\textbf{Statement of Disclosure:} The authors report there are no competing interests to declare. 
%\end{keyword}
\par
\,

\section{Introduction and main idea} \label{Introduction}

The intersection of artificial intelligence, sentiment analysis, and financial markets has been a fertile ground for research over the last decade. As the ability of models to understand and classify human sentiments has improved, so has the allure of using these capabilities to predict stock price movements. In this paper, we harness the power of GPT-4 to evaluate the sentiments of financial tweets and examine their effect on same-day stock price movements for Apple and Tesla. We additionally compare the results to sentiments derived by BERT.

Microblogging, exemplified by platforms like Stocktwits or Twitter, is a digital communication form where users share short, frequent updates. Thus, it allows for the rapid dissemination of information and sentiment, often faster than traditional news outlets. Events like the GameStop stock surge of 2021 illustrated how microblogging communities can rally around and influence stock prices, demonstrating the power of collective sentiment in real-time financial decisions.

Sentiment analysis can be applied to a variety of tasks, such as asessing customer feedback, analysing reactions to breaking news or perceptions of social media content. It can provide insights into people's feelings about news related to stock companies and can help determine how such news is received by the public, whether it's perceived as positive or negative. These perceptions might be linked to stock price movements. Given the high frequency of news and social media content in this digital age, the importance of computer-driven sentiment analysis grows. Computers can process content instantly, which is crucial for identifying relevant content both quickly and accurately. As such, enhancing sentiment analysis techniques to detect nuances like sarcasm and irony through context-based analysis becomes even more vital.

In our study, we assume that a higher correlation between investor sentiment and same-day stock return movements indicates a better interpretation of the social media feed. Therefore, we assume that finding a stronger predictability between same-day price movements and the sentiment of social media feed suggests that the utilized sentiment model (GPT-4 vs BERT) is more effective in capturing investor sentiment. To focus our study solely on the performance of modeling microblogging sentiment towards stock movements, we remove the potentially diluting effect of an uncertain link between past sentiments and future returns. Sentiment derived from tweets might possess a limited shelf-life, implying that today's hot topics or sentiments may not necessarily influence the stock's movement tomorrow or the day after. By predicting same-day movements, we capture the most immediate and potent effects of sentiment.

%reasonung behind same-day prediction
Despite the abundance of research articles on forecasting specific returns, it remains evident that two schools of thought prevail. One asserts the possibility of gaining abnormal returns through certain investment strategies, while the other contends that the efficient market hypothesis is fundamentally true, resulting in such returns being unsustainable over the long term.
Supporting our approach, the Efficient Market Hypothesis (EMH) (\cite{fama1970efficient}) posits that stock prices rapidly adjust to new information, making it challenging to predict future price movements based solely on historical data. Consequently, modeling same-day reactions is more feasible than attempting to forecast longer-term changes.
Given that we have employed large and closely monitored companies like Apple and Tesla, we assume that the markets are quite efficient in these domains. Thus, detecting a possible link between future returns and current sentiments would likely require further steps in these specific contexts. However, should such a link be established—meaning for stocks lacking efficient markets—the methodology we have developed here would be adequately applicable. It would aid in exploring these patterns more comprehensively.
To reiterate, our main intention is to demonstrate the potential for a connection between sentiment and capital stock price movement, contingent upon an intrinsic link between the two factors and showing how investors can capture this link using a sophisticated prompt.

Following this introduction, section \ref{sec_review} provides an overview of the relevant literature. In section \ref{modelsetup}, we detail the model setup, our methodology, message handling, prompt engineering, and our approach to stock data and sentiment matching. Section \ref{result} presents the results of our analysis, highlighting key findings. Finally, in section \ref{discussion}, we engage in a thorough discussion of our results and conclude the article with major takeaways and potential future research directions.

\section{Review of the literature} \label{sec_review}

Sentiment analysis in stock price prediction has been explored through many approaches and techniques. To offer a clear comparative perspective, Table~\ref{tab:literature_overview} encapsulates key details from the highlighted studies. It summarizes the data sources, time spans, sentiment methods, performance metrics, and prediction models utilized by various authors in the recent years. This tabulation provides an organized reference, emphasizing the diversity and nuances in methodologies adopted by researchers in this field.

The field of sentiment analysis has seen significant advancements in recent years, with researchers exploring various models and methodologies to understand and predict stock market movements. The adoption of sophisticated models for extracting investor sentiment is a trend that has piqued the interest of numerous scholars. A noteworthy contribution in this context is the work by \cite{leippold2023sentiment}, which emphasized the significance of using more advanced sentiment models. Specifically, they showcased how sentences manipulated using GPT-3 can yield semantically incoherent results that are, however, easily recognized by humans. By instructing GPT-3 to generate synonyms for negative words and then using these to rephrase sentences, they evaluated the robustness of models. They underscore FinBERT's resilience against adversarial attacks, especially when compared to traditional keyword-based methods.

While there's a surge in studies employing cutting-edge models for sentiment analysis, there's no shortage of holistic overviews and thorough reviews that trace the evolution of this field over time. \cite{wankhade2022survey} provided an overview of sentiment analysis, offering a comprehensive review of the subject, its methodologies, applications, and developments in the field.
The review of \cite{rodriguez2023review} encompassed both traditional methods and newer models, including BERT and GPT-2/3, spotlighting their roles and advancements in the domain of sentiment analysis.

Within the ambit of sentiment analysis, GPT models, particularly the more recent GPT variants, have emerged as favorites in contemporary studies.
\cite{kheiri2023exploiting} employed the GPT-3.5 Turbo model to undertake sentiment analysis on social media posts. For comparative analysis and benchmarking, the authors utilized RoBERTa. Their study attributed the specific role social scientist to the model.
\cite{belal2023leveraging} also used the GPT 3.5 Turbo variant, for their sentiment analysis on Amazon reviews. They revealed a major enhancement in accuracy. As benchmark models for sentiment classification, they used VADER and TextBlob. However, the specifics of the prompt they employed remained unspecified in their publication. \cite{lopez2023can} employed a range of AI models, including GPT-1, GPT-2, GPT-3, GPT-4, and BERT, to predict stock market returns based on news headlines. Interestingly, they found that GPT-1, GPT-2, and BERT models are not particularly effective in accurately predicting returns. Further, they used a regression model to predict future returns. Meanwhile, \cite{zhang2023unveiling} predicted Chinese stock price movements using sentiment analysis, employing models such as ChatGPT, Erlangshen RoBERTa, and Chinese FinBERT.

Besides more recent work, earlier works have also explored the correlation between public mood and stock market movements.
\cite{bollen2011twitter} analyzed 9.9 million tweets to uncover a correlation between the Dow Jones Industrial Average and public mood. They employed OpinionFinder to determine daily positive vs. negative sentiments and Google-Profile of Mood States to assess moods in six dimensions. Their predictive model utilized granger causality analysis and neural networks.
Build upon Bollen's work, \cite{mittal2012stock} crafted a unique mood analysis based on four classes (calm, happy, alert, kind) derived from the POMS questionnaire. Alongside the granger causality test, they integrated machine learning models like SVM and neural networks for enhanced prediction accuracy.
\cite{rao2012analyzing} demonstrated a significant correlation between stock prices and Twitter sentiment, including DJIA, NASDAQ-100, and 13 other major technology stocks. In addition to assessing a "bullishness" metric sentiment, they measured an agreement measure among positive and negative tweets. Their findings revealed a positive correlation between stock returns and concurrent day bullishness.
\cite{oliveira2017impact} utilized Twitter data to predict returns, volatility, and trading volume of indices and portfolios, analyzing 31 million tweets related to 3,800 US stocks. Their methodology incorporated rolling window and regression models. For sentiment analysis, they employed a lexicon-based approach. Similar to \cite{rao2012analyzing}, they computed a Bullish/Bearish Ratio, an agreement ratio, and a variation measure to capture the nuances in public sentiment.
\cite{ranco2015effects} examined Twitter sentiments concerning 30 stock companies from the DJIA. Their study revealed a low Granger causality and person correlation between stock prices and Twitter sentiments. To determine the sentiment, 100,000 tweets were manually annotated, with 6,000 of them annotated twice to gauge expert consensus. Support Vector Machines (SVM) were utilized as the primary classification method.
\cite{coqueret2020stock} looks at the link between company sentiment and future returns and does not find news sentiment to be a powerful predictor.
\cite{hamraoui2022impact} explored the relationship between Twitter sentiments and the Tunisian financial market. The study found a low Pearson correlation and Granger causality between the two.
\cite{matthies2023to} employed Twitter data to investigate its correlation with subsequent day stock volatility. For this, they used linear regression models and did not uncover a significant relationship between sentiment, as measured by the VADER sentiment analysis tool, and stock performance. Their dataset comprised approximately 2 million tweets associated with four stocks known for high price volatility. Notably, their findings underscored that sentiment typically did not enhance the predictive power of their models significantly.
In contrast \cite{smailovic2013predictive} indicated that sentiment trends could predict price changes days ahead. They calculated a daily "positive sentiment ratio" by comparing the number of positive tweets to the total tweets, and looked at how this ratio changed daily.
\cite{si2013exploiting} leveraged topic-based sentiments from Twitter to forecast the stock market. Aiming for a one-day-ahead prediction of the stock index, they employed a Vector Autoregression (VAR) model for predictions. Their approach started with topic modeling using the Dirichlet Processes Mixture (DPM) to determine the number of topics from daily tweets. Once topics were identified, they crafted a sentiment time series based on these topics. Finally, both the stock index and the sentiment time series were integrated into an autoregressive framework for predictions.
\cite{bing2014public} explored the potential of public sentiment in predicting the stock price of specific companies. Their study, spanning 30 companies from NASDAQ and NYSE, analyzed 15 million tweets.
\cite{pagolu2016sentiment} investigated the correlation between stock price movements of a company and the public opinions expressed in tweets about that same company. To analyze sentiment, they employed techniques such as word2vec and n-gram models.
\cite{jaggi2021text} introduced the FinALBERT and ALBERT models in their examination of stocktwits. They categorized stocktwits based on the same-day stock price changes and used various machine learning techniques to predict these categories. 

Some inquiries have shifted their focus from merely the content to the credibility and influence of certain Twitter personalities or financial circles in sentiment analysis. This underscores not just the essence of the content but also the significance of its originator. \cite{yang2015twitter} focused on identifying relevant Twitter users within the financial community. The study then found a correlation between a weighted sentiment measure using messages from these essential users and major financial market indices.
\cite{gross2019buzzwords} explored the connection between Twitter sentiment and stock returns. They focused on expert users whose tweets predominantly revolved around financial topics. For sentiment analysis, they utilized a dictionary-based approach.
\cite{sul2017trading} looked at the feelings expressed in 2.5 million Twitter posts about individual S\&P 500 companies and compared this to the stock market performance of those companies. They found that tweets from users with fewer than 171 followers (which was the median) that weren't retweeted had a big effect on a company's stock performance the next day, and also 10 and 20 days later. They used the Harvard-IV dictionary to analyze the sentiments in the tweets.

Also the timeframe of the sentiment analysis is an important factor. Recent global events, such as pandemics, have also influenced the direction of research.
\cite{valle2022does} studied the relationship between Twitter sentiment and the performance of financial indices during pandemics, specifically focusing on H1N1 and Corona. They considered both fundamental and technical indicators. To gauge sentiment, they used a lexicon-based approach, and their analysis was centered on tweets from financial-focused Twitter accounts.
\cite{katsafados2023twitter} examined tweets related to the COVID pandemic, utilizing the VADER sentiment analysis approach. They measured the degree of positivity and negativity in tweets and linked this information to the stock market. Their results indicated that, in the short term, heightened positivity is linked to increased returns, whereas negativity was associated with diminished returns.

In addition to sentiment analysis, some studies have introduced other methods to consolidate news for stock market forecasts.
\cite{jiang2021pervasive} introduced an innovative approach to aggregating news. They constructed portfolios based on news-driven and non-news driven returns, buying when news returns were high, selling when low, and holding for 5 days.
\cite{bustos2020stock} presented a comprehensive overview of stock movement prediction. They highlighted that most papers in their study used market information or technical indicators as input variables, with some also incorporating news, blogs, or social network data.
\cite{sousa2019bert} employed BERT for sentiment analysis of news articles and predicted future DJI index movements. They also manually labelled news data as positive, neutral, or negative, and fine-tuned the BERT model. Additionally, they looked at the hours of sentiment before opening time and predicted subsequent DJI day trends.

Lastly, the integration of various data sources, including social media texts and technical indicators, has been a focal point for many researchers.
\cite{ji2021stock} incorporated both social media texts and technical indicators into their predictive model. They utilized daily stock price data. Rather than deriving a sentiment measure from the textual content, they employed Doc2Vec to extract text features. To effectively process multiple posts from a single day, they aggregated the data into daily groupings.

\newpage

\begin{landscape}
	\begin{table}[htbp]
		\centering\caption{Structured Literature Overview}
		\label{tab:literature_overview}
		\scalebox{0.8}{
			\begin{tabular}{ |p{4cm}||p{1.6cm}|p{3.5cm}|p{5.5cm}|p{6cm}|p{5.5cm}|  }
				%\hline
				%\multicolumn{8}{|c|}{Literature Overview} \\
				\hline
				Article	&	Data source	&	Time-span	&	Sentiment method	&	Performance Measure	&	Model/Method	\\
				\hline \hline
				\cite{bing2014public}	&	Twitter	&	Oct 2011 - Mar 2012		&	Dictionary-based	&	Accuracy	&	Own Data Mining Algorithm, Naive Bayes, SVM	\\ \hline
				\cite{bollen2011twitter}	&	Twitter	&	Feb 2008 - Dec 2008	&	OpinionFinder, Google Profile	&	Accuracy, MAPE	&	Granger Causality, Neural Networks	\\ \hline
				\cite{coqueret2020stock}	&	Twitter, News	&	May 2012 - Nov 2017	&	Bloomberg Sentiment Score	&	T-statistic, Rsquared	&	Regression (linear)	\\ \hline
				\cite{gross2019buzzwords}	&	Twitter	&	Jan 2010 - Sep 2018	& 	Dictionary-based	&	Accuracy, ROC AUC, DM test	&	SVM	\\ \hline	
				\cite{hamraoui2022impact}	&	Twitter	&	Oct 2016 - Sep 2017	&	Semantic Analysis	&	VAR, T-statistic, Pearson correlation	&	Granger Causality 	\\ \hline
				\cite{jaggi2021text}	&	Stocktwits	&	2010 - 2020	&	FinALBERT, ALBERT	&	Accuracy, F1	& Regression (logistic), Random Forest, Naive Bayes, Gradient Boosting	\\ \hline
				\cite{ji2021stock}	&	Comments, News	&	Jan 2010 - Nov 2019	&	Doc2Vec	&	MAE, RMSE, R-squared	&	LSTM neural network	\\ \hline
				\cite{katsafados2023twitter}	&	Twitter	&	Jan 2021 - Jun 2021	&	VADER	&	Pedroni's cointegration test (p-values)	&	Panel ARDL/VAR Regression	\\ \hline
				\cite{lopez2023can}	&	News	&	Oct 2021 - Dec 2022	&	GPT-1, GPT-2, GPT-3, GPT-4, BERT	& Accuracy, F1, Regression Coefficients, R-squared, AIC, BIC	&	Regression (linear and logistic)	\\ \hline
				\cite{matthies2023to}	&	Twitter	&	Oct 2020 - Jun 2021	&	VADER	&	R-squared	&	Regression (linear)	\\ \hline
				\cite{mittal2012stock}	&	Twitter	&	Jun 2009 - Dec 2009	&	POMS questionaire	&	Accuracy	& Regression (linear and logistic), SVM, Neural Network (SOFNN)	\\ \hline
				\cite{oliveira2017impact}	&	Twitter	&	Dec 2012 - Oct  2015	&	Lexicon-based	&	DM test for predictive Accuracy	&	Multiple Regression, Neural Network, SVM, Random Forest	\\ \hline
				\cite{pagolu2016sentiment}	&	Twitter	&	Aug  2015 - Aug 2016	&	Word2vec, n-grams	&	Accuracy, ROC, Precision, Recall, F-Measure	&	Regression (logistic), LibSVM	\\ \hline
				\cite{ranco2015effects}	&	Twitter	&	Jun 2013 - Sep 2014	&	Manual annotation, SVM	&	significance of increase in CAR values	& Event study, Granger Causality, Pearson Correlation	\\ \hline
				\cite{rao2012analyzing}	&	Twitter	&	Jun 2010 - Jul 2011	&	Naive Bayesian Classification, API Twittersentiment	&	MaxAPE, R-squared	&	Granger Causality, Regression (linear), EMMS	\\ \hline
				\cite{si2013exploiting}	&	Twitter	&	Nov 2012 - Feb 2013	&	DPM	&	Accuracy	&	VAR model	\\ \hline
				\cite{smailovic2013predictive}	&	Twitter	&	Mar 2011 - Dec 2011	&	Manually labelled and SVM	&	Accuracy	&	Granger Causality	\\ \hline
				\cite{sousa2019bert}	&	News	&	May 2018 - Feb 2019	&	Manual annotation, Fine-tuned BERT, SVM, TextCNN, Naive Bayes	&	Accuracy, Precision, Recall, F1	&	Comparison between DJI variation and overall sentiment	\\ \hline
				\cite{sul2017trading}	&	Twitter	&	Mar 2011 - Jan 2013	&	Harvard-IV dictionary	&	Regression analysis (linear)	&	Regression (linear)	\\ \hline
				\cite{valle2022does}	&	Twitter	&	Jun - Jul 2009; Jan - May 2020	&	Lexicon-based (S140, AFFIN, SenticNet)	&	Correlation	&	Correlation between sentiment and stock market (shifted)	\\ \hline
				\cite{yang2015twitter}	&	Twitter	&	Feb 2014 - Jun 2014	&	SentiWordNet (dictionary approach)	&	Regression analysis (linear)	&	Regression (linear)	\\ \hline
				\cite{zhang2023unveiling}	&	News	&	Oct 2021 - Feb 2023	&	ChatGPT, Erlanshen RoBERTa, Chinese FinBERT	&	Annual excess return, Annual net asset return, Win rate, Sharpe Ratio 	&	Portfolio	\\ \hline
				\hline
		\end{tabular}}
	\end{table}
\end{landscape}

\newpage

\section{Model setup} \label{modelsetup}

\subsection{BERT and GPT}
The development of deep learning models, particularly the Transformer architecture, has marked a significant shift in the way textual data is processed. At the heart of this transformation lie two influential models, GPT (Generative Pre-trained Transformer) and BERT (Bidirectional Encoder Representations from Transformers). These models harness the power of the Transformer architecture, developed Google AI, introduced in the paper of \cite{vaswani2017attention}.
The main objective of the Transformer model was to enhance and outperform existing text processing paradigms. Traditional models often processed text word-by-word, leading to limitations. Transformers, on the other hand, allowed entire sentences to be input, thus enabling not only parallelization, but also consider the context of words. The concept of self-attention was also introduced by \cite{vaswani2017attention}, which enabled the model to process words in the context of surrounding words. In their original Transformer model, the architecture is divided into an encoder and a decoder. However, variations of the Transformer architecture might use only one of these components or both, depending on the specific application. For example, BERT leverages the encoder (\cite{devlin2018bert}), while GPT uses a multi-layer decoder (\cite{radford2018improving}).
BERT has two predominant models: BERT-Base, composed of 12 encoder blocks and BERT-Large, using 24 encoder blocks (\cite{devlin2018bert}).
During its pre-training phase, BERT is exposed to the prediction of masked tokens and next sentence predictions. This training harnesses data from BooksCorpus (comprising 800M words) and English text passages from Wikipedia (encompassing 2,500M words). \cite{devlin2018bert} showcased the model's adaptability by fine-tuning BERT on eleven NLP tasks.

GPT was first introduced by OpenAI in 2018 with GPT-1 and has since seen multiple iterations, each improving upon the other in terms of model size, data, and tasks they can handle. It has been trained on a diverse range of datasets including Wikipedia articles, books, and internet data.
The architectural divergence between BERT and GPT is primarily based on directionality. BERT operates bidirectionally, it comprehends a word by analyzing its preceding and succeeding context. GPT, in contrast, is uni-directional, primarily forecasting a word based on preceding words. Conversely, for our GPT sentiment analysis, we engage the GPT-4 API with a customized prompt, which we will describe in section \ref{prompt} of this study. GPT-3 (\cite{brown2020language}) uses the same model architecture as GPT-2 (\cite{radford2018improving}). 
GPT-4, released in 2023, is a successor to GPT-3 and build on the same foundational ideas. GPT-4, like its predecessors, is a generative model from OpenAI built upon the Transformer architecture. However, its distinguishing features lie in its sheer scale, improved fine-tuning capabilities, and broader application potential. This model excels in few-shot learning, adapting to specialized tasks with minimal examples. With a broader training dataset and potential architectural tweaks, GPT-4 sets a new standard in the domain of natural language processing.
Further information about GPT-4 can be found in the technical report on the OpenAI website.

\subsection{Data and Methodology}
To investigate the potential improvement of the GPT-4 LLM in comparison to BERT, we have decided to study the daily stock price movements of the companies Apple and Tesla for the year 2017\footnote{We have checked the validity of our approach using BERT also for years from 2016 to 2020 and results were similar. However, as GPT-4 API requests are costly and 2017 had the least number of Stocktwits, we decided to use this year as our year of study.}. For that, we have recorded adjusted closing prices $C_t$ for each day $t$ and company of the study from the yahoo Finance platform. These are translated into up-down movements:
\begin{equation}\label{eq1}
U_t = 
\begin{cases} 
1 & \text{if } C_t > C_{t-1} \\
0 & \text{otherwise}
\end{cases}
\end{equation}
This results in overall $T=502$ combined daily observations, of which Apple ($App$) contributes $T_{App}=251$ and Tesla ($Tes$) $T_{Tes}=251$ observations. To create a sentiment for each day of these observations, we have used the micro blogging website Stocktwits. Stocktwits is a social media platform designed for investors, traders, and individuals interested in the stock market and financial markets in general. It functions as a microblogging platform where users can share their thoughts, opinions, and insights about various stocks, cryptocurrencies, and other financial instruments. Users post short messages, often referred to as ``tweets'', containing stock tickers, charts, news articles, and commentary on market trends. These messages are public and can include text, images, and links. The Stocktwits messages have been obtained from the researcher team of \cite{jaggi2021text} to ensure comparability between research in the field.
The selected Stocktwit messages of Apple and Tesla, have been juxtaposed to the stock price data to match the prices' dates. For the whole time period of the year 2017, we have extracted 214,548 messages which were the foundation for the scheduled sentiment analysis. 
Each of these messages have then been analyzed to extract a sentiment using the two LLMs. 
\subsubsection{Message handling} \label{msg_hand}
As both LLMs have different capabilities, we needed to clean the messages differently for each model. In case of the GPT-4 model, we have removed all URL or other references to websites, removed duplicate entries of tweets (as they are most likely advertisement), removed all images and reformatted every tweet to lowercase. To remove duplicates we have used the full data range of \cite{jaggi2021text} starting from 2010-06-28, so we can more efficiently see potential advertisements which are assumed to have no relevant sentiment. The BERT model needed further cleaning, as we have observed that certain elements prevent it from producing a plausible sentiment. Therefore, in addition to the cleaning steps for GPT-4, we have also removed all hashtags (\# -lead words), cashtags (\$ -lead words), mentions (@ -lead words), plain unicodes (e.g. emoticons) and numbers (e.g. dates, percentages) as well as all special characters (e.g. . and ;) for BERT only.
In the message box \ref{box1} you can find an exemplary Stocktwit compared with the formatting for each model.
\begin{figure}[ht]
\centering
\begin{mybox}[Formatting differences]
\textbf{Original:} \$AAPL OK, bought \$162.50 calls, my shares sitting fine from forever ago...LONG \\
\textbf{GPT:} \$aapl ok, bought \$162.50 calls, my shares sitting fine from forever ago...long \\
\textbf{BERT:} ok bought calls my shares sitting fine from forever ago long \\
Timestamp: 2017-10-18 13:51:24 UTC 

\end{mybox}
\end{figure}
To extract the sentiment for BERT, we have used the pretrained weights for BERT called ``bert-base-multilingual-uncased-sentiment'' from NLP town, downloadable at www.huggingface.co. This specific variant of BERT was trained on product reviews which had ratings of 1 star (negative) to 5 stars (positive) and is therefore specifically tailor made for sentiment analysis. We have not made any further changes to the model. As a result of classifying all messages using these model and weights, we receive the logits for each sentiment, which can be recomputed to give us probabilities for each class from 1 to 5. The class with the highest probability is consequently chosen as the final sentiment classification. 
For GPT-4, we have used the API model version of August 2023 in combination with the 8k token environment. To remove unnecessary creativity for the LLM's answers and gain reproducibility of results, we have set the model parameter ``temperature'' to 0. However, as GPT is not tailor-made to produce sentiments from messages, we needed to fine-tune the user prompts to come up with a comparable classification, while at the same time utilizing the powers of the LLM beyond simple sentiment analysis.
\subsubsection{Prompt engineering} \label{prompt}
In order to access the full capabilities of an LLM, a new field called ``prompt engineering'' has emerged. The term refers to the practice of designing and refining input prompts (e.g. commands) to guide the output of mostly language models, in order to achieve desired results or improve the model's performance on specific tasks. It involves the careful crafting of questions, instructions, or other inputs to maximize the quality and relevance of the model's responses. Determining the correct prompts is therefore essential for sentiment analysis using LLMs. On the one hand, we need the machine to produce concise and comparable statements and on the other hand, we need to allow the machine to use its contextual based capabilities to provide high-quality outputs. The former is relevant as each request to such a machine is costly, in terms of CPU/GPU-time and money. Each unnecessary input or output token will therefore increase the processing time as well as the money paid and consequently might render the original purpose of the analysis pointless. Eventually the results of each prompt need to be summarized in some way, e.g. by averaging, counting, semantic analysis or other. Therefore, the otherwise highly appreciated variety in answers is usually not desired in terms of sentiment analysis. Saving resources by truncating the prompt too much, however, does lead to issues with the quality of the output, which in terms of sentiment analysis comes mostly from the skill to analyze news based on context. Message box \ref{box2} outlines this issue. 

\begin{figure}[ht]
\centering
\begin{mybox}[Prompt differences]
\textbf{Original:} the president will lose the fight with apple manufacturing \$aapl \\
\textbf{Sentiment of an imprecise prompt:} '1(neg)': 0.7, '2': 0.15, '3': 0.05, '4': 0.05, '5(pos)': 0.05 \\
\textbf{Sentiment of our prompt:} '1(neg)': 0.2, '2': 0.2, '3': 0.2, '4': 0.2, '5(pos)': 0.2

\end{mybox}
\end{figure}

In the original tweet, we can clearly see a negative general sentiment, indicated by words like ``lose'' and ``fight''. In terms of the results of the imprecise prompt, we can clearly see that this is captured quite well, as the negative sentiment 1 has the highest probability with 70 percent. However, from the perspective of the company Apple this message does not convey bad news. In fact, this should be perceived as good news, as the president losing is a win for Apple in this case. Our prompt can capture this ambiguity much clearer by assigning each sentiment the same probability of 20 percent. Further note that both prompts were designed to explicitly provide the sentiment probabilities without further text, fulfilling the requirements of a concise and comparable prompt. In code box \ref{code1} we therefore show the our final prompt, which has been sent to the API. \footnote{The imprecise prompt is created by using our own prompt as base, but then removing all requests on advantage and relatedness as well as changing the system string to ``You are a financial analyst, who analyzes sentiment for Apple (AAPL).''}

\begin{lstlisting}[style=pythonstyle, caption={Python code to create our prompt.}, label=code1]
sentiment = ['1(neg)','2','3','4','5(pos)']
relation = ['Mostly Apple','Mostly competitor','Unrelated']
advantage = ['Advantage','Disadvantage']


def sentiment_function(Message):
    openai.api_key = "KEY" 
    response =openai.ChatCompletion.create(
        model="gpt-4",
        messages=[
            {"role": "system", "content": """You are a financial analyst, who analyzes if news could have some benefits for the company Apple (AAPL)."""},
            {"role": "user", "content": f"""
            Determine from {sentiment}: probabilities, from {advantage}: probabilities, from {relation}: probabilities.
            Format: [Sentiment: Probabilities for each sentiment, Advantage: Probabilities, Relation: Probabilities]. 
            Alternatively, state "NA".   
            ```{Message}```"""
             },

        ],
        temperature = 0

    )
    ans = response['choices'][0]['message']['content']
    return ans
\end{lstlisting}

Our prompt has several features, which we have found to be relevant for a correct sentiment. These features guarantee the three before mentioned requirements of producing concise, comparable and contextual results. These are:

\begin{itemize}
\item \textbf{Providing a role for GPT-4:} It is of crucial importance, to allow the GPT-4 LLM to assess the situation from a specific perspective. For that, specifying a role like ``financial analyst'' can aid in the process to achieve better contextual results. (Line 11)
\item \textbf{Set a subject and goal:} The GPT-4 LLM can benefit from further context, if a specific situation or setting is described, which allows for more tailor-made answers. For that, setting an aim for the analysis can tremendously help with the outcome. Here, we have set that potential benefits for a company need to be evaluated. The subject of interest, e.g. Apple, must be defined clearly, especially if multiple words for expressing the subject (e.g. Apple and AAPL) exist. (Line 11)
\item \textbf{Add ``catch-all'' elements:} Even if we are interested in the sentiment alone, it is very beneficial to add further perspectives on the matter. Keep in mind that many of the microblogging tweets will have totally unrelated information or could be sarcastic. Therefore, allowing the LLM to think of the sentiment in terms of advantages for or relatedness to the company improves the prompt even further. (Lines 1,2,3,14)
\item \textbf{Extracting probabilities:} Probabilities help the model and the user to understand, that some decision are easier than others and that there is very often room for interpretation. These probabilities, which are usually expressed as logits in other LLMs, help clarify these uncertainties. (Line 13)
\item \textbf{Using lists:} To shrink the output and therefore the number of tokens required, it is very useful to command GPT-4 to use a specific format. Here, the python syntax of [] for a list helps. This also allows for better comparison of results. (Line 14)
\item \textbf{Specify an alternative:} If all efforts are in vain, then GPT needs an alternative. Otherwise, the LLM will very likely produce a text body explaining why it could not come up with an answer. (Line 15)
\item \textbf{Zero shot prompt:} Zero shot prompts will use one prompt to immediately access the power of the GPT-4 LLM to achieve a result. This helps to reduce cost, but can lead to deterioration of quality in comparison to few shot prompts, where the machine can see some exemplary results. 
\end{itemize}

Besides that, we have also used specific formatting using e.g. three consecutive double quotes for multi line strings. To see a proper guidance for formatting, we would like to refer to the OpenAI course for GPT on the deeplearning.ai platform.

Each of the 214,548 Stocktwit messages have been handled according to section \ref{msg_hand} and then sent to GPT-4 using the prompt described here. The prompt has been adjusted for Tesla by simply replacing the words Apple and AAPL with Tesla and TSLA. As each Stocktwit was of different character length and in about 15-20 percent of cases the returned answer was simply "NA", requests exhibited varying costs. On average, however, we concluded that each prompt was roughly 0.01 USD.

\subsubsection{Stock data and sentiment matching} \label{match}

The resulting microblogging sentiment probabilities $P^S$ do not follow an equidistant time gap between messages and are of higher frequency than daily and therefore need to be matched properly to be used for explaining daily price movements. We have therefore decided to aggregate the data using averages as well as counting. We apply the following procedure for each of the companies separately. 

For the sentiments, we have classified a message based on the probability, whereas the highest probability was determining the final sentiment $s$:

\begin{equation*}
s=\argmax_{j}(P^S_{ij}) \quad \forall i \in \{1, 2, ..., m\},
\end{equation*}

where \( j \) ranges over all columns of \( P^S \) and $m$ is the number of all messages. As can be obtained from code box \ref{code1}, we have sentiments as integers from 1 (negative) to 5 (positive), i.e. $j \in \{1,2,3,4,5\}$. We have then calculated the average of the sentiments $s$ for a time window of 16:00 ET  of $t-1$ to 16:00 ET of $t$ for every day $t$ in the study and deducted 3 to receive 0 as the baseline case:

\begin{equation}\label{eq2}
\bar{s}_t = \left( \frac{\sum_{i=1}^{N} s_i}{N} \right) - 3,
\end{equation}

where $N$ is the number of messages for the given time window. For the GPT-4 decision of ``advantage'', we again have classified the presumed advantage of the message for the company by taking the $m$ provided probabilities $p_a$ of the GPT-4 answer. We then calculate the simple sum of occurrences for advantage $a_t$ and disadvantage $d_t$ during the same time frame as the sentiment $\bar{s}_t$:\footnote{The possible regressor for ``Relation'' was investigated but dropped, as it did not have a positive and significant impact on the outcome.}

\begin{eqnarray*}
a_t &=&\sum_{i=1}^{N} \bm{1}_{\{p_{a,i} > 0.5\}} \\
d_t &=&\sum_{i=1}^{N} \bm{1}_{\{p_{a,i} < 0.5\}}
\end{eqnarray*}

This allows us to retrieve sentiments $s$, advantages $a$ and disadvantages $d$ as regressors to match with the stock price movements $U_t$ from equation \eqref{eq1} for each company, having exactly $T_{App}$ and $T_{Tes}$ observations respectively. We have also constructed the average advantage $\bar{a}_t$ equivalently to \eqref{eq2}, but will only use this object for the exploratory data analysis and not for the model, as the combined objects $a_t$ and $d_t$ also inherit relevant information of the number of messages per day.

For the final modeling part, we now need to stack these regressors to be able to conduct a regression on the full sample with $T=502$ observations and allowing for differences in parameters based on the company. Stacking regressors is a common procedure to isolate group specific effects, see e.g. \cite{otto2023estimation} for a similar application. We apply this for the sentiments $\bar{s}_{App,t}$ of Apple and $\bar{s}_{Tes,t}$ of Tesla as follows:

\begin{align}\label{eq3}
\bm{s}_{App} &= \begin{bmatrix}
0 \\
0 \\
\vdots \\
0 \\
\bar{s}_{App,1} \\
\bar{s}_{App,2} \\
\vdots \\
\bar{s}_{App,T_{App}}
\end{bmatrix}
&
\bm{s}_{Tes} &= \begin{bmatrix}
\bar{s}_{Tes,1} \\
\bar{s}_{Tes,2} \\
\vdots \\
\bar{s}_{Tes,T_{Tes}} \\
0 \\
0 \\
\vdots \\
0
\end{bmatrix}
\end{align}

where there are $T_{App}=T_{Tes}$ zero entries per vector.  The same procedure is applied for $a_t$ and $d_t$ for each company, resulting in $\bm{a}_{App}$, $\bm{a}_{Tes}$ and $\bm{d}_{App}$, $\bm{d}_{Tes}$.

Please note, that this setup allows us to model the upward or downward movement of Apple and Tesla solely based on the sentiment and contextual analysis of Stocktwit messages of that same day. Detecting a link between the captured context of the messages with GPT-4 and the price movements would allow us to infer that this model is indeed sufficient to be applied in a financial context. We therefore clearly want to point out that we do not follow the goal of proving that there is an exploitable link of microblogging messages towards future returns. All following results have been obtained using the programming language R, depictions are created using the package DescTools \citep{desctools2017}.

\section{Results} \label{result}
\begin{figure}[ht]
    \centering

    \includegraphics[width=0.99\linewidth, scale=1.0]{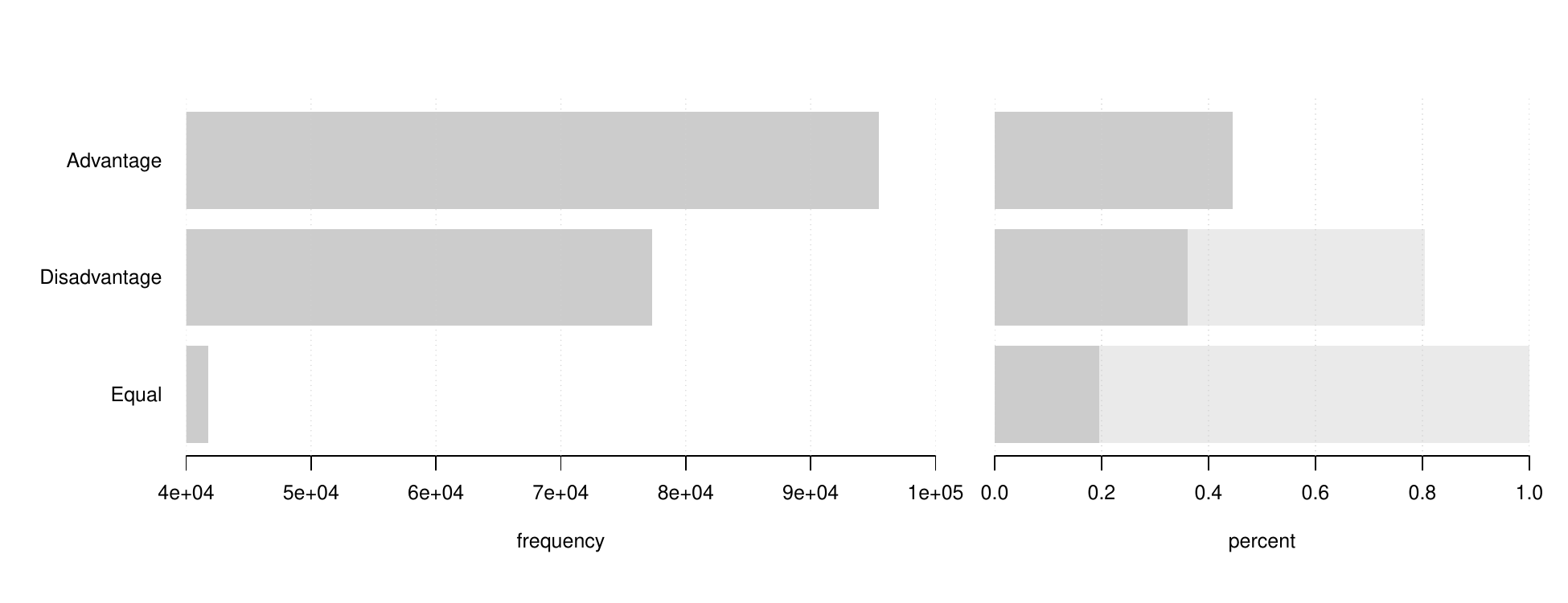} % Adjust 
    
    \caption{Absolute and relative counts for the estimation results of a message being advantageous or disadvantageous. Equal is used when $p_{a,i} = 0.5$.}
    \label{fig1}
\end{figure}

First and foremost, we have analyzed the outcome of our prompts engineering approach described in sections \ref{prompt} and \ref{match} respectively. For that, we are specifically interested in the addition of the advantages estimations of each message. Figure \ref{fig1} provides an overview of the different classifications by GPT-4.

Here it can be obtained, that during our study in 2017, most micro bloggers of Stocktwit have been publishing complaisant information towards the companies. This is reflected by the fact, that almost 50 percent of messages have been classified as of being advantageous for either Apple or Tesla. In twenty percent of cases, however, GPT-4 could not decide whether a message favors the company or not, therefore specifying the probability for advantage and disadvantage as being 0.5. Message box \ref{box3} provides more insights for this case:\footnote{Probabilities of BERT are rounded and thus might not sum up to 1.}

\begin{figure}[ht]
\centering
\begin{mybox}[Prompt differences]
\textbf{Adjusted Message:} stock option trades screener 4 trade ideas \$vxx \$ea \$wmt \$aapl wmt \$bip \$info \$vale \$aig \$vrx \$tbt \$nvda \$x \$oled	 \\
\textbf{GPT evaluation:} [Sentiment: '1(neg)': 0.1, '2': 0.1, '3': 0.2, '4': 0.3, '5(pos)': 0.3, Advantage: 'Advantage': 0.5, 'Disadvantage': 0.5, Relation: 'Mostly Apple': 0.1, 'Mostly competitor': 0.7, 'Unrelated': 0.2] \\
\textbf{Bert evaluation:} '1': 0.10, '2':	0.12, '3':	0.25, '4':	0.29, '5':	0.23

\end{mybox}
\end{figure}

In this box we can clearly see, that the message is not displaying any clear favor for the mentioned companies, as it not clear what ``trade idea'' is being proposed - it could be buying or selling the stock. GPT-4 catches this by specifying the same probability for advantage as for disadvantage. BERT, on the other hand, which is not capable of analyzing the context, will simply lean towards a positive sentiment, without understanding that trade ideas alone could be good or bad for the company.

It is furthermore relevant, to understand if the proposed usage of our regressor ``Advantage'' is not simply the same as a positive sentiment, as this would render all efforts of a more sophisticated prompt useless. We therefore have decided to specifically analyze the univariate and bivariate relationship of both, by using the previously mentioned average sentiment $\bar{s}_t$ and average advantage $\bar{a}_t$. Keep in mind that the average sentiment has a support on $[-2,2]$ by construction and the average advantage on $[-1,1]$, while 0 being neither positive nor negative or advantageous or disadvantageous respectively. In addition, we analyze how the grouping variable company effects the distribution of sentiment and advantage, as we specifically constructed company based regressors such as the ones seen in \eqref{eq3}. Figure \ref{fig2} visualizes our analysis for that:

\begin{figure}[ht]
    \centering
    \begin{subfigure}{0.48\textwidth}
        \centering
        \includegraphics[width=\linewidth, height=0.15\textheight, scale=1.0]{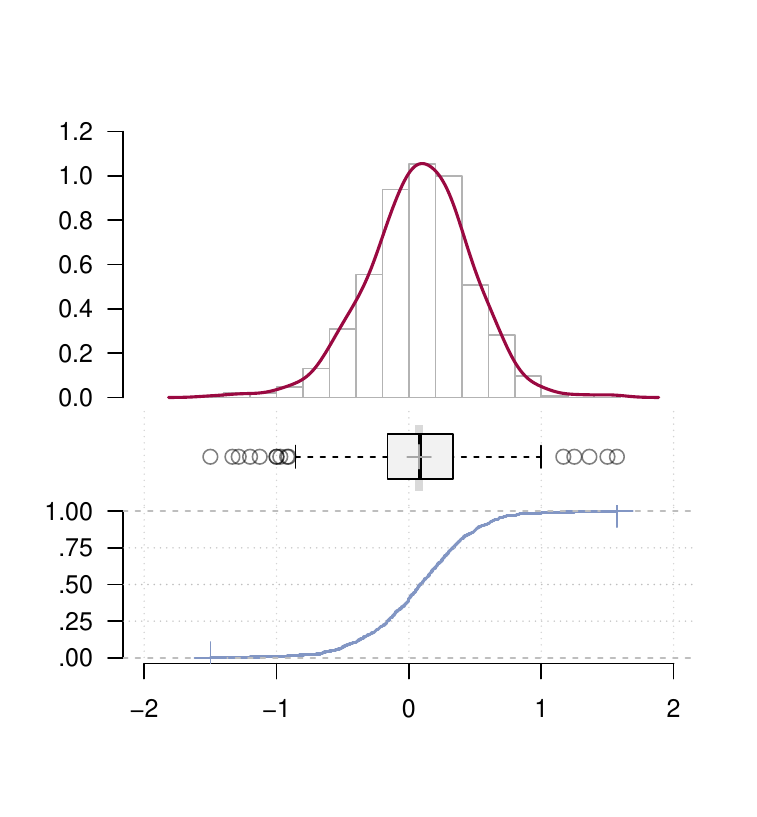}
        \caption{Distribution of average sentiment $\bar{s}_t$}
    \end{subfigure}
    \hfill
    \begin{subfigure}{0.48\textwidth}
        \centering
        \includegraphics[width=\linewidth, height=0.15\textheight, scale=1.0]{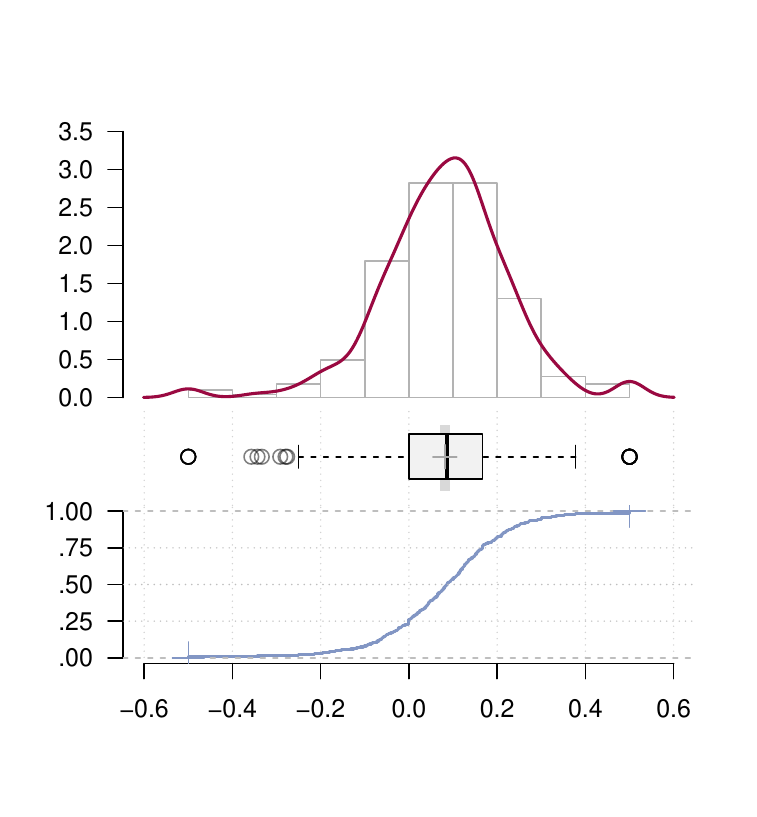}
        \caption{Distribution of average advantage $\bar{a}_t$}
    \end{subfigure}

    \vspace{1em} % Some vertical spacing between the two rows of images
    
    \begin{subfigure}{\textwidth}
        \centering
        \includegraphics[width=0.99\linewidth, scale=1.0]{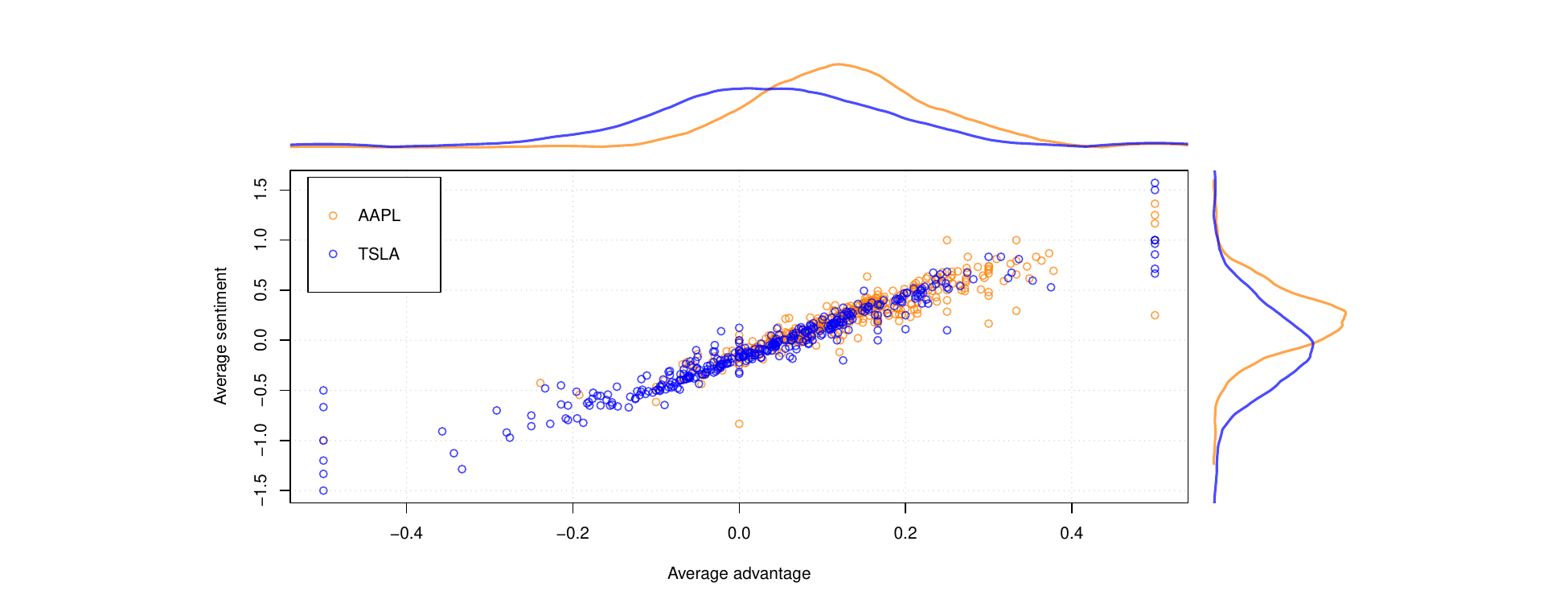} % Adjust the width as needed
        \caption{Average sentiment ($\bar{s}_t$) with average advantage ($\bar{a}_t$) by company}
    \end{subfigure}
    
    \caption{Univariate and marginal densities for the average sentiment and the average advantage.}
    \label{fig2}
\end{figure}

This figure, containing 3 subplots, provides various relevant information. First and foremost, we can obtain from the subfigures (a) and (b) that the distribution of sentiment and advantage is seemingly smooth, with some heavy tails for the average advantage, i.e. some days for which we find very strong implications for an advantage or disadvantage. This in accordance with stock price movements when they are modeled as returns. Here, it is well known that these returns are exhibiting heavy tails, see e.g. \citep{cont2001empirical}. Besides that, we can obtain from the bivariate scatterplot, that sentiment and advantage do have a clear and strong correlation, which is to be expected. In most cases, messages directed to the companies' Stocktwit feed are presented to determine an advantage or disadvantage for a company by using a certain tone of the message. Nonetheless, there are many dots observable in the scatterplot, which show a discrepancy of the general seemingly linear relation of sentiment and advantage. Message box \ref{box4} provides insights into this:

\begin{figure}[ht]
\centering
\begin{mybox}[Prompt differences]
\textbf{Adjusted Message:} \$vhc +8.6\% (reportedly being awarded \$344 mln from apple (\$aapl dispute)	 \\
\textbf{GPT evaluation:} [Sentiment: '1(neg)': 0.1, '2': 0.1, '3': 0.1, '4': 0.1, '5(pos)': 0.6, Advantage: 'Advantage': 0.1, 'Disadvantage': 0.9, Relation: 'Mostly Apple': 0.7, 'Mostly competitor': 0.2, 'Unrelated': 0.1] \\
\textbf{Bert evaluation:} 
'1':0.59, '2':	0.22, '3':	0.12, '4':	0.04, '5':	0.02

\end{mybox}
\end{figure}

Analyzing this message, it is clear that this is not good news for Apple. However, as there are words like ``awarded" and ``+8.6'' GPT-4 is deciding on a positive sentiment. However, it can due to its contextual skills, understand that this message is disadvantageous for Apple as there seems to be a payment for VHC being due. Interestingly, BERT does capture the negative sentiment, very likely because of the word ``disaster'' and the fact that for BERT numbers like ``+8.6'' have been removed. Further analyzing Figure \ref{fig2} also helps to understand, why it is necessary to treat companies individually in any further modeling part. For the data of 2017 it is clear that there is a great favoritism for Apple, indicated by the mode of both distributions being in the positive range. For Tesla, the mode is around 0 for both, sentiment and advantage. We therefore conclude that our adjustments in equation \eqref{eq3} are justified.

Other interesting insights from the contextual capabilities of GPT-4 emerge from the fact that is also can use the background knowledge it was trained on. this becomes particularly noteworthy when it comes to finance based terminology, as shown in message box \ref{box5}.

\begin{figure}[ht]
\centering
\begin{mybox}[Prompt differences]
\textbf{Adjusted Message:} \$aapl ok, bought \$162.50 calls, my shares sitting fine from forever ago...long	 \\
\textbf{GPT evaluation:} [Sentiment: '1(neg)': 0.1, '2': 0.1, '3': 0.1, '4': 0.2, '5(pos)': 0.5, Advantage: 'Advantage': 0.7, 'Disadvantage': 0.3, Relation: 'Mostly Apple': 0.9, 'Mostly competitor': 0.05, 'Unrelated': 0.05] \\
\textbf{Bert evaluation:} 
'1': 0.01,     '2': 0.03,   '3': 0.32, '4':      0.41, '5':    0.24
\end{mybox}
\end{figure}

This message states that the users is going ``long'' and that the users bought ``calls'' which is finance terminology for buying an asset or betting on increasing stock prices respectively. Paired together with the word ``fine'' GPT-4 can easily deduct that this is not only advantageous for Apple, but also indicative of a high positive emotion towards the stock. This can not be done using simpler models like BERT, which do not have the background knowledge. Thus, BERT only classifies this message as slightly positive.\footnote{We also have evidence from messages like ``\$nvda \$tsla this is samazing news for \$nvda'' where the probability for that message concerning a competitor was estimated to be very high by GPT-4. However, later experiments by us showed that incorporating this into the model does not yield better results, so we decided to not provide further insights into that feature of our prompt.}	

Finally, we have investigated the capabilities of both, BERT and GPT-4 LLM, for modeling stock price movements solely based on microblogging sentiment. For that, we have chosen a logistic regression approach using several training and test splits.\footnote{Throughout our analysis, we have employed also other models like xgboost or deep neural networks. Given the relatively few number of daily observations in combination with the number of regressors, we have found that logistic regression performs best in terms of model accuracy.} Logistic regression is a statistical method used for analyzing datasets in which there are one or more independent variables that determine an outcome. In our case, the outcome is the movement of stock prices, which can be categorized as either "up" or "down". By using the logistic regression model, we can predict the likelihood of a stock price moving up based on our created regressors.
For logistic regression, the log-likelihood function \( \ell(\beta) \) for a given set of coefficients \( \beta \) is defined as:

\[
\ell(\beta) = \sum_{i=1}^{T} \left[ U_i \log(p_i) + (1-U_i) \log(1-p_i) \right],
\]

where \( T \) is the number of observations in the dataset, \( U_i \) is the observed stock price movement of Apple or Tesla for the \( i^{th} \) observation and \( p_i \) is the predicted probability of the \( i^{th} \) observation being 1, given by the logistic function:
\[
p_i = \frac{1}{1 + e^{-(\beta_0 + \beta_1 X_{1,i} + \beta_2 X_{2,i} + \ldots + \beta_k X_{k,i})}}
\]

The objective in logistic regression is to find the values of \( \beta \) that maximize this log-likelihood function. We use Fisher Scoring to obtain the coefficient estimates. For BERT, we only have $k_{BERT}=2$ regressors, the average stacked sentiments $\bm{s}_{App}^{BERT}$ for Apple and $\bm{s}_{Tes}^{BERT}$ for Tesla. For GPT-4 we have $k_{GPT}=6$ regressors, the average sentiments $\bm{s}_{App}^{GPT}$ and $\bm{s}_{Tes}^{GPT}$ as well as the counts for stacked advantage and disadvantage $\bm{a}_{App}^{GPT}$, $\bm{a}_{Tes}^{GPT}$, $\bm{d}_{App}^{GPT}$ and $\bm{d}_{Tes}^{GPT}$, see \eqref{eq3} for an explanation. To maintain a possible relationship in time of the daily stock price movements, we split our data into training and test observations using connected periods of calendar months. We have conducted several splits, where the training period always starts with January 2017 and ends in different months. The test period starts with the upcoming month and always ends in December 2017. This is done to provide information on the stability of results and to overcome issues with potential seasonal effects in the data. We measure the success of a model in terms of accuracy. Accuracy is a metric used to evaluate classification models. It represents the proportion of correct predictions in the total predictions made. The formula for accuracy in terms of True Positives (TP), True Negatives (TN), False Positives (FP), and False Negatives (FN) is given by:

\[
\text{Accuracy} = \frac{TP + TN}{TP + TN + FP + FN}.
\]

As a benchmark model we use a simple buy-and-hold strategy and refer to it as ``Naive''. Additionaly to the accuracy comparison, we have investigated the p-values of the bivariate paired series of BERT with the Naive model logits of the test data and the GPT with Naive model logits using McNemar's test. The results of our experimental setup can be obtained from Table \ref{tab1}.

\begin{table}[h]
    \centering
    
    \begin{tabular}{ccccccc}
        \toprule
        \textbf{Begin test} & \textbf{Naive} & \textbf{BERT} & \textbf{$p_{Bert | Naive}$} & \textbf{GPT-4} & \textbf{$p_{GPT | Naive}$} & \textbf{$n_{test}$} \\
        \midrule
        April & 51.85\% & 65.87\% & \num{2.29e-4} & 70.11\% & \num{3.85e-5} & 378 \\
        May & 52.06\% & 66.47\% & \num{9.10e-3} & 71.47\% & \num{4.91e-5} & 340 \\
        June & 51.69\% & 65.54\% & \num{9.48e-2} & 69.26\% & \num{1.42e-2} & 296 \\
        July & 51.59\% & 65.87\% & \num{1.91e-4} & 66.27\% & \num{1.06e-3} & 252 \\
        August & 50.47\% & 66.51\% & \num{4.87e-2} & 66.51\% & \num{8.58e-4} & 212 \\
        September & 46.99\% & 65.66\% & \num{5.23e-2} & 66.87\% & \num{6.02e-4} & 166 \\
        \bottomrule
    \end{tabular}
    \caption{Comparison of accuracies for stock price movements: BERT vs. GPT-4. Number of observations in the test sample shown in $n_{test}$. All p-values are calculated using the naive buy-and-hold strategy as benchmark.}
    \label{tab1}
\end{table}

Across the evaluation period, the Naive baseline displayed a fluctuating accuracy, ranging from 46.99\% to 52.06\%. BERT's performance consistently surpassed this, with its best accuracy observed at 66.47\% during the month of May. In terms of statistical significance, BERT's p-values remained notably low, underscoring the reliability of its predictions, particularly in April, May, and July where the p-values dropped below 0.001, indicating that it is able to reflect the sentiment of microblogging sentiment towards stock price movements.

GPT-4, on the other hand, consistently demonstrated a commendable performance, outpacing BERT in five out of the six months. Its peak accuracy reached 71.47\% in May, with the lowest p-value observed in the same month at $\num{4.91e-5}$. All p-values are below the common threshold of 0.05. This suggests a consistent and statistically significant predictive power of the GPT-4 model in the context of stock price sentiment analysis.\footnote{We have excluded test periods beginning in January, February, March, October, November and December as for all of these cases either the training or the test period would have only three months for estimation and application respectively. This would most likely lead to unstable and not representative results.}

\section{Discussion and Conclusion} \label{discussion}
The application of GPT-4 in stock market sentiment analysis offers a novel approach to understanding market dynamics and investor sentiment. Our study primarily focused on the sentiments of Apple and Tesla stock from various microblogging messages over a year. The study underscores the power and potential of GPT-4 in analyzing microblogging sentiments for financial applications. By refining input prompts, GPT-4 was able to capture nuanced sentiments and context that might be missed by other models like BERT. The comparison between the two models highlighted the advanced contextual understanding capabilities of GPT-4, making it a promising tool for sentiment analysis in finance.

However, while our findings suggest a strong correlation between sentiment derived from GPT-4 and stock price movements, there are several considerations and implications to be discussed.

While traditional models for sentiment analysis, such as BERT and its variants, have been widely used, GPT-4's ability to understand context and nuance might offer a more accurate representation of sentiment. Its performance in our study supports this claim, but further comparative analyses are needed to establish its superiority conclusively. This is specificall true as the significance of prompt engineering in obtaining desired outputs from GPT-4 cannot be understated. The quality and specificity of the prompts play a pivotal role in sentiment extraction. Future studies can explore the optimal strategies for prompt engineering tailored for financial sentiment analysis.

The use of GPT-4 introduces a new dimension in sentiment analysis, but its cost might be prohibitive for some investors or analysts. However, the potential benefits in terms of accuracy and insight generation might outweigh the costs, especially for institutional investors.

Beyond sentiment analysis, there's potential to use GPT-4 and similar models in other areas of financial analysis, such as forecasting, risk assessment, and portfolio optimization. Stock prices can be influenced by a myriad of factors beyond just microblogging sentiments, which we have not explored within this study. The versatility of GPT-4 opens avenues for multifaceted financial research.

The insights derived from this study can also be beneficial for hedge funds, institutional investors, and individual traders in making informed decisions. Moreover, news agencies and financial platforms might integrate such sentiment analysis tools to offer real-time sentiment scores alongside stock prices, providing an additional layer of information for their audience.

In conclusion, while the GPT-4 LLM offers promising insights and capabilities in the realm of financial sentiment analysis, further studies, and refinements are essential for its practical applications in modeling stock price movements.

\bibliographystyle{apalike} 
\bibliography{gpt_bib} 

\clearpage

\end{document}